\documentclass[preprint,showpacs]{revtex4}
\usepackage{epsfig}
\usepackage{graphicx}
\usepackage{bm}
\newcommand{\bmath}[1]{\mbox{\boldmath{${#1}$}}}

\newcommand{\half}{\mbox{${\textstyle \frac{1}{2}}$}}           

\newcommand{\rd}{\textrm{d}}
\begin{document}
\title{Pion Compton scattering and bremsstrahlung}
\author{G\"oran F\"aldt}\email{goran.faldt@tsl.uu.se} 
\author{Ulla Tengblad}\email{ulla.tengblad@tsl.uu.se} 
\affiliation{ Department of nuclear and particle physics, 
Uppsala University,
 Box 535, S-751 21 Uppsala,Sweden }
%

%
\begin{abstract}
The pion-polarizability functions are structure functions 
of pion-Compton scattering. They can be assessed in high-energy 
pion-nucleus bremsstrahlung reactions, 
$\pi^- +A\rightarrow\pi^- +\gamma +A$.
We present numerical expectations for pion-nucleus 
bremsstrahlung cross sections in the Coulomb region, 
{\it i.e.}~the small-angle region where the nuclear scattering 
is dominated by the Coulomb interaction. We investigate 
the prospects of measuring the polarizability functions for 
pion-Compton c.m.\ energies
from threshold up to 1 GeV. A meson-exchange model is used 
for the pion-Compton amplitude.
\end{abstract}
\maketitle
%
%
%
\section{Introduction}

The cross-section distribution for high-energy pionic bremsstrahlung
\[
\pi^- +A\rightarrow\pi^- +\gamma +A  ,
\]
is at small-momentum transfers to the nucleus dominated by the 
one-photon-exchange contribution.
Hence, by measuring  cross-section distributions for 
pion-nucleus bremsstrahlung at small momentum
transfers one gains at the same time information
about the cross-section distribution for pion-Compton scattering
\[
\gamma+\pi^- \rightarrow \gamma+\pi^-;
\]
the photon of the initial state being the virtual photon exchanged 
between pion and nucleus. In particular, it becomes 
possible to determine pion polarizabilities.

The theoretical underpinnings of the pion-bremsstrahlung reaction
were investigated by  Gal'perin et al.\ \cite{4}, and an 
experiment  was susequently performed by  Antipov  et al.\ \cite{Ant},
which showed that pion polarizabilities could indeed be determined.
The COMPASS experiment at CERN \cite{COMP} is a refinement of the Antipov
experiment.

In ref.\cite{FT} a mathematical expression for the bremsstrahlung cross
section was derived assuming a pion-Compton amplitude that in additon 
to the Born contributions contained also contributions from the
 $\sigma$, $\rho$, and $a_1$, exchanges. The kinematics of the nuclear
 reaction was defined through
\begin{equation}
\pi^-(p_1)+ {\rm A}(p)\rightarrow \pi^-(p_2)+\gamma(q_2)+{\rm} A(p'),
\end{equation}
and the kinematics of the related pion-Compton reaction through
 \begin{equation}
\pi^-(p_1)+ \gamma(q_1)\rightarrow \pi^-(p_2)+\gamma(q_2) , \label{Compt-kin}
\end{equation}
 with $q_1=p-p'$. In terms of these kinematic variables the cross-section 
 distribution can be written as \cite{FT}
\begin{eqnarray}
\frac{\rd \sigma}{\rd^2q_{1\bot}  \rd^2q_{2\bot} \rd x}
  &= &\frac{4Z^2\alpha^3}{\pi^2m_{\pi}^4}
 \Bigg[ \frac{\bmath{q}_{1\bot}^2}{(\bmath{q}_{1\bot}^2+q_{min}^2)^2}\Bigg]
  \Bigg[ \frac{1-x}{x^3} \Bigg] 
   \Bigg[ \left( \frac{x^2 m_{\pi}^2}{\mathbf{q}_{2\bot}^2+x^2 m_{\pi}^2}\right)^2\Bigg] 
         \cdot   \nonumber \\     \nonumber \\    
   &&\Bigg[ \left| A(x,\mathbf{q}_{2\bot}^2)\right|^2
    \left\{ 1 - \mu^2\frac{4 x^2 m_{\pi}^2\mathbf{q}_{2\bot}^2}
                          {(x^2 m_{\pi}^2 +\mathbf{q}_{2\bot}^2)^2} \right\} \nonumber \\  
                             \nonumber \\   
    && \qquad+2\Re\left(A(x,\mathbf{q}_{2\bot}^2)
        B^{\star}(x,\mathbf{q}_{2\bot}^2)\right)
    \left\{ 1 - \mu^2\frac{2 \mathbf{q}_{2\bot}^2}
                          {x^2 m_{\pi}^2 +\mathbf{q}_{2\bot}^2} \right\}\nonumber \\  
    && \qquad+ \left|B(x,\mathbf{q}_{2\bot}^2)\right|^2\Bigg] .
 \label{Coul-peak-cross-distr}
 \end{eqnarray}
 This expression is valid when the transverse momenta of the 
 emerging pion and photon are much smaller than their longitudinal
 momenta.
The parameter $\alpha$ is the fine-structure constant, 
$m_{\pi}$ the pion mass, $x$ the ratio 
\begin{equation}
x=
  \frac{q_{2z}}{p_1} =\frac{\omega_2}{E_1} ,  \label{x-fraction}
\end{equation} 
$\mu=\hat{\mathbf{q}}_{1\bot}\cdot\hat{\mathbf{q}}_{2\bot}$, 
and $q_{min}$ the minimum-longitudinal-momentum transfer  
\begin{equation}
q_{min}
	  =\frac{m_{\pi}^2}{2E_1}\cdot \frac{x}{1-x}  .\label{qmin-def}
\end{equation}
At high energies  $q_{min}$  is  extremely small. 

The functions 
$A(x,\mathbf{q}_{2\bot}^2)$ and $B(x,\mathbf{q}_{2\bot}^2)$ characterize
the pion-Comp\-ton amplitude. In the Born approximation
\begin{eqnarray}
	A(x,\mathbf{q}_{2\bot}^2)=1 , \label{ABorn}\\
	B(x,\mathbf{q}_{2\bot}^2)=0  ; \label{BBorn}
\end{eqnarray}
and generally  
\begin{eqnarray}
	A(x, \mathbf{q}_{2\bot}^2)&=&  1 -\frac{x^2}{4(1-x)}
	\left( \frac{\mathbf{q}_{2\bot}^2+x^2 m_{\pi}^2}{x^2 m_{\pi}^2}\right)^2
	   \lambda_1(x, \mathbf{q}_{2\bot}^2)  , \label{Atild-fin} \\
  B(x, \mathbf{q}_{2\bot}^2)&=& 
   \frac{x^2}{2(1-x)}\left(\frac{\mathbf{q}_{2\bot}^2+x^2 m_{\pi}^2}{x^2 m_{\pi}^2}\right)
   \lambda_2(x,\mathbf{q}_{2\bot}^2)  ,\label{Btild-fin} 
\end{eqnarray}
with $\lambda_1(x, \mathbf{q}_{2\bot}^2)$ and $\lambda_2(x, \mathbf{q}_{2\bot}^2)$
generalized-polarizability functions.
Their mathematical expressions, in the meson-exchange model,
are given in \cite{FT}.

The  structure of the cross-section distribution is mainly determined by
the inelastic-Coulomb factor
\begin{equation}
\frac{\bmath{q}_{1\bot}^2}{(\bmath{q}_{1\bot}^2+q_{min}^2)^2} \label{Prim_fact}
\end{equation}
which vanishes when the transverse-momentum transfer $\bmath{q}_{1\bot}$
to the nucleus vanishes. This is the classic Primakoff factor, which
exhibits  a maximum at
${q}_{1\bot}=q_{min}$. When the momentum transfer ${q}_{1\bot}$ 
becomes much larger than $q_{min}$ the Coulomb contribution to the 
bremsstrahlung cross section becomes
small and the hadronic  contribution dominates \cite{PrimakI}. This
contribution is not considered in the present paper.

The Mandelstam kinematic variables of the pion-Compton scattering (\ref{Compt-kin}) 
can in our application be approximated as follows;
\begin{equation}
\begin{array}{rcl}
  s - m_{\pi}^2 &=& \displaystyle{ \frac{1}{x(1-x)} }
    \bigg[\mathbf{q}_{2\bot}^2 + x^2 m_{\pi}^2 \bigg]  ,\\ \\
  t &=&  \displaystyle{\frac{-1}{1-x} }\bigg[\mathbf{q}_{2\bot}^2 + x^2 m_{\pi}^2 \bigg]  ,\\ \\
  u - m_{\pi}^2&=&  \displaystyle{\frac{-1}{x} }\bigg[\mathbf{q}_{2\bot}^2 + x^2 m_{\pi}^2 \bigg]  .
\end{array}    
\label{Mandel}
\end{equation}  
\newpage
\section{Kinematical considerations}

We are interested in studying the pion-Compton-scattering process. 
In order to do so in a transparent way we need to make specific cuts 
in the pion-nucleus-bremsstrahlung phase space. To this end 
a few kinematical observations might be helpful.

Let us first introduce, instead of the pion-Compton 
c.m.~energy and the photon-transverse momentum, 
the corresponding dimensionless variables
\begin{eqnarray}
	w   & = & \sqrt{s}/m_{\pi} \\
	y & = & |\mathbf{q}_{2\bot}| /	m_{\pi}  ,
\end{eqnarray}
with $w\geq 1$ and $y\geq 0$.
The first entry of Eq.(\ref{Mandel}), defining the
 pion-Compton c.m.~energy squared, can then be reformulated as 
\begin{equation}
		y^2=x[(1-x)w^2-1]  . \label{q2-def}
\end{equation}
Since $y^2$ is positive we conclude that
for a given fixed value of $w$ the variable  $x$ must 
be in the interval 
\begin{equation}
	0\leq x \leq x_{max}=1 -\frac{1}{w^2} . \label{x-ineq}
\end{equation}
The maximal value of $y$ occurs at the midpoint of
this interval, 
\begin{equation}
	y_{\rm max}=\frac{w}{2}(1 -\frac{1}{w^2}) . \label{Y-max-point}
\end{equation}
We may also rewrite Eq.(\ref{q2-def}) as
\begin{equation}
	y^2=x(x_{max}-x) w^2 , \label{x-max-transverse}
\end{equation}
showing directly that the photon-transverse momentum vanishes
at the kinematic end-points $x=0$ and $x=x_{max}$.

For a fixed value of $x$, on the other hand, it follows from
the positivity of $y^2$ of Eq.( \ref{q2-def}) that the 
pion-Compton c.m.~energy is bounded from below
\begin{equation}
	w^2 \geq 1/(1-x).  \label{w-ineq}
\end{equation}
The boundary value obtains in the forward direction, where 
the photon-transverse momentum vanishes.

In Fig.1   the photon-transverse momentum $y(x,w)$ is plotted for a few 
values of the pion-Compton c.m.~energy $w=\sqrt{s}/m_{\pi}$. 
As an illustration  consider $\sqrt{s}=2m_{\pi}$, a mass 
at which  the threshold approximation to the 
polarizability functions should be acceptable. In this case 
the maximal value 
of $x$ is $0.75$ and the maximal value of the 
photon-transverse momentum $q_{2\bot}$ is $3m_{\pi}/4$. 
\begin{figure}[ht]\begin{center}
\scalebox{0.40}{\includegraphics{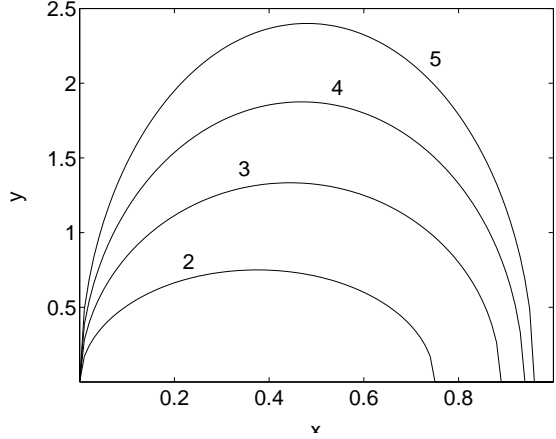}}
\end{center}
\caption{Photon-transverse momentum $y=q_{2\bot}/m_{\pi}$ as a function
of $x$ for fixed values of the Compton mass	$w=\sqrt{s}/m_{\pi}=2,$ 3, 4, 5. 
The threshold value, $w=1$, corresponds to the point $y=x=0$.}
\end{figure}
\clearpage
\section{Low-mass-Compton region} 

Hard-pion bremsstrahlung offers a unique possibility of determining 
pion-thresh\-old polarizabilities. However, this will succeed only
if we restrain the c.m.\ mass in the pion-photon system to masses 
well  below the $\rho$-meson mass, so as to avoid the
$s$-channel $\rho$-exchange contribution to the Compton amplitude. 

Hence, we first concentrate our efforts
on the region $s\ll m_{\rho}^2$, 
where we can  confidently approximate 
the structure functions 
 $A(x,\mathbf{q}_{2\bot}^2)$ and $B(x,\mathbf{q}_{2\bot}^2)$ 
of Eqs.(\ref{Atild-fin}) and (\ref{Btild-fin}) by their near-threshold values, with
\begin{eqnarray}
\lambda_1(x, \mathbf{q}_{2\bot}^2)&=&\frac{m_{\pi}^3}{\alpha}\ 
  (\alpha_{\pi} +\beta_{\pi})  =0.0006      , \label{Lamb1-thr} \\
     \lambda_2(x,\mathbf{q}_{2\bot}^2)&=&\frac{m_{\pi}^3}{\alpha}\ \beta_{\pi} 
       = -0.0131 ,\label{Lamb2-thr} 
\end{eqnarray}
and $\alpha_{\pi}$ and $\beta_{\pi}$  the standard pion-electric and 
pion-magnetic polarizabilities. These values given
are obtained in the meson-exchange model when fixing the $\sigma$-meson 
coupling so that the chiral Lagrangian value \cite{Hol} of $\lambda_2$ is
reproduced. Our present knowledge of the pion polarizabilities is
reviewed in \cite{Ahrens}.

The cross-section distribution (\ref{Coul-peak-cross-distr}) contains 
angular dependent terms, through 
$\mu=\hat{\mathbf{q}}_{1\bot}\cdot\hat{\mathbf{q}}_{2\bot}$. 
Measured variables are however 
$\mathbf{q}_{2\bot}$ and $\mathbf{p}_{2\bot}$, but since, 
$\mathbf{p}_{2\bot}=\mathbf{q}_{1\bot}-\mathbf{q}_{2\bot}$, with  
${q}_{1\bot}\ll {q}_{2\bot}$, the angular variations of 
$\mathbf{q}_{2\bot}$ and $\mathbf{q}_{1\bot}$
are independent. Thus, the integrations over 
the angles of $\mathbf{q}_{2\bot}$ and $\mathbf{q}_{1\bot}$
in Eq.(\ref{Coul-peak-cross-distr}) are unrestricted 
and we may replace  $\mu^2$ by its angular average, which is 
a half, and at the same make the replacement
\begin{equation}
\rd^2q_{1\bot}  \rd^2q_{2\bot}=\pi^2\rd q_{1\bot}^2  \rd q_{2\bot}^2  .
\end{equation}

After integration over angles the cross-section distribution factorizes
into factors that depend either on ${q}_{1\bot}$ and $x$, or on 
${q}_{2\bot}$ and $x$. The integration over the 
inelastic-Coulomb factor (\ref{Prim_fact}) is an 
integration over ${q}_{1\bot}$,
but there is an $x$ dependence in $q_{min}$,
\begin{equation}
q_{min}
	  =\frac{m_{\pi}^2}{2E_1}\cdot \frac{x}{1-x}.
\end{equation}
It is tempting to integrate over all possible momentum transfers 
${q}_{1\bot}$, but that is not possible since the integral diverges.
The reason is that we have neglected pion and nuclear 
form factors. Also, for large values of ${q}_{1\bot}$ the 
hadronic-bremsstrahlung contribution dominates  the Coulomb
contribution. Consequently, we must make a cut so as to avoid the hadronic
contribution. The best approach is to make a cut in the same way for
all $x$ values. This is achieved by cutting off the integral
at $r_- q_{min}$ on the low-momentum side and at $r_+ q_{min}$ on
the high-momentum side of the Coulomb peak. From  (\ref{Prim_fact})
we derive 
\begin{equation}
P_f\equiv P(r_- , r_+) =\int_{r_- q_{min}}^{r_+ q_{min}}\rd {q}_{1\bot}^2
\frac{\bmath{q}_{1\bot}^2}{(\bmath{q}_{1\bot}^2+q_{min}^2)^2}  .
          \label{Peak_fact}
\end{equation}
This integral is obviously independent of $x$. The precise values
of $r_-$ and $r_+$ to be chosen can only be decided after 
inspection of the experimental distribution.

Next we integrate the cross-section distribution 
Eq.(\ref{Coul-peak-cross-distr}) over  transverse-photon momenta 
in the domain
	 $ 0\leq q_{2\bot}^2 \leq q_{2\bot max}^2$. Since the 
	 polarizability functions $\lambda_1$ and $\lambda_2$
entering the structure functions $A(x,\mathbf{q}_{2\bot}^2)$ and $B(x,\mathbf{q}_{2\bot}^2)$ of Eqs.(\ref{Atild-fin}) and 
(\ref{Btild-fin}) are now constants the integration is
straightforward, and we get
\begin{equation}
 \frac{\rd \sigma(x, q_{2\bot max})}{ \rd x}
  = \frac{4Z^2\alpha^3}{m_{\pi}^2}
    \cdot \frac{1-x}{x}   \ 
      P_f   F(z)    ,
 \label{Lin-cross-distr}
 \end{equation}
with
\begin{equation}
	z(x,q_{2\bot max})=\frac{q_{2\bot max}^2}{x^2m_{\pi}^2}
	  \label{z-def}
\end{equation}
and the distribution function 
\begin{equation}
	F(z)= u(z)  +
       \frac{2x^2}{1-x} \left( -\frac{1}{4}v(z)\lambda_1
             +\frac{1}{2}w(z)\lambda_2\right)   . \label{F-funct}
\end{equation}
The $x$-dependent  prefactor in Eq.(\ref{Lin-cross-distr}) is simply 
the ratio of  the final state momenta, $(1-x)/x =E_2/\omega_2$, and
  the singularity at  $x\approx0$ is the well-known 
soft-photon-radiation singularity.

\begin{figure}[ht]\begin{center}
\scalebox{0.35}{\includegraphics{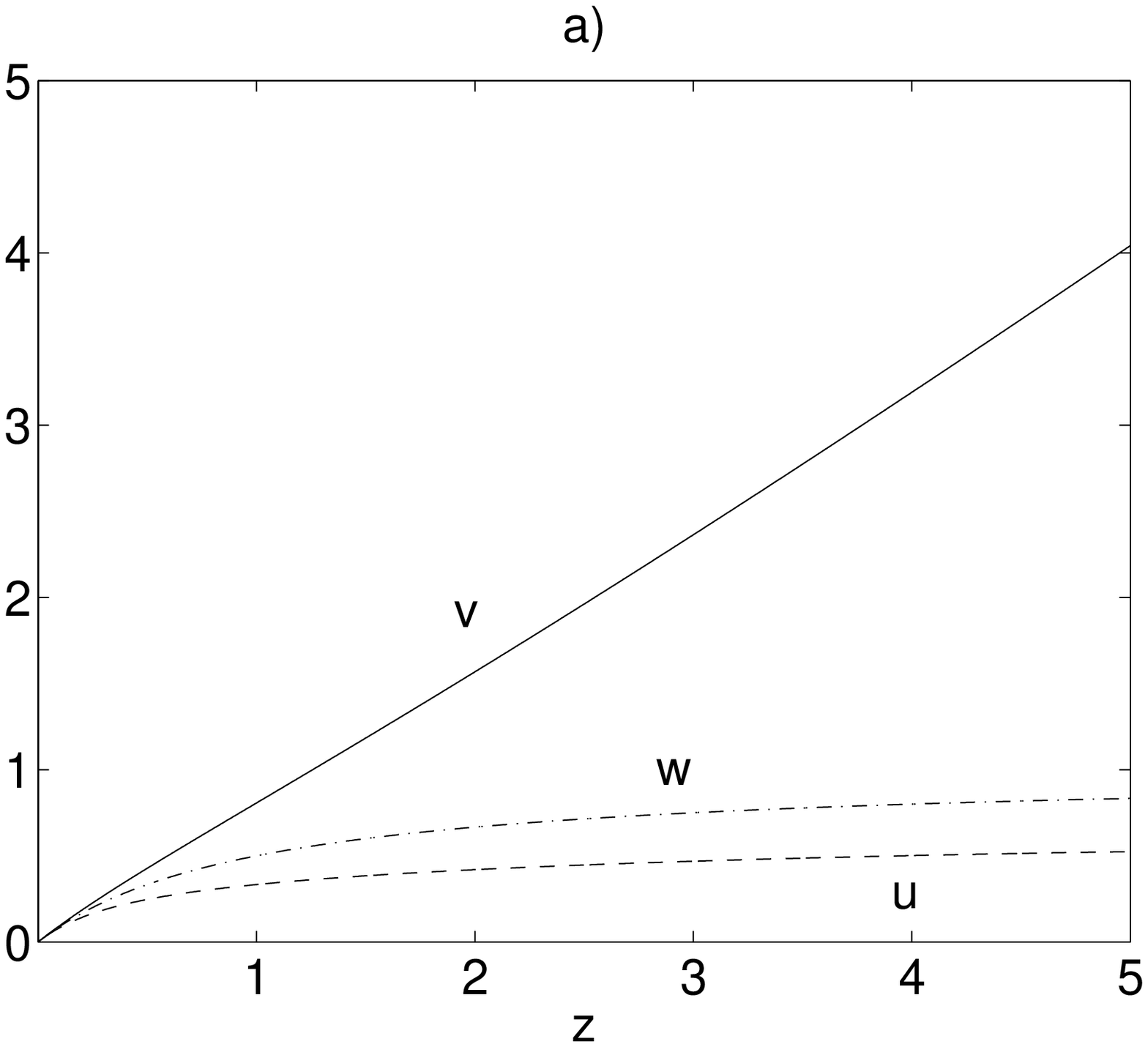}
 \qquad \quad \quad \includegraphics{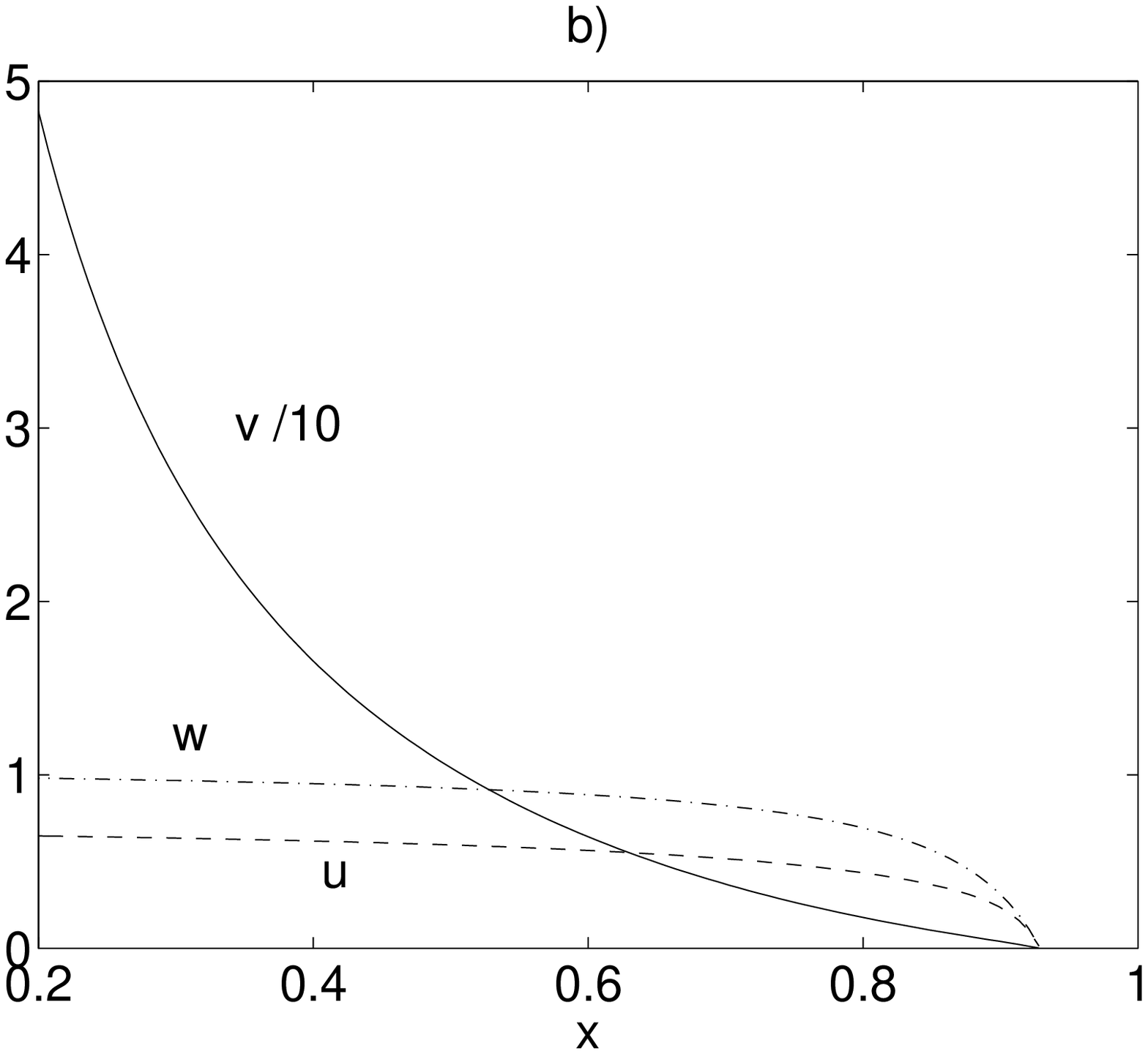}}
\end{center}
\caption{Plots of $u$, $v$, and $w$ as 
a) functions of $z$; and b) functions of $x$ for 
$\sqrt{s_{max}}=3.75 m_{\pi}$. In b) we have plotted $v/10$ instead of $v$.}
\end{figure}

The functions $u(z)$, $v(z)$, and $w(z)$, of Eq.(\ref{F-funct}) 
are all elementary and given in the Appendix. They are graphed in Fig.2a.
For small values of the argument $z$,
or  transverse momenta such that 
$q_{2\bot max}^2\ll x^2 m_{\pi}^2$, we have
\begin{equation}
	u(z) \approx v(z) \approx  w(z) \approx  z  .\label{thesh-appr}
\end{equation}
For large values of the argument $z$,
or  transverse momenta such that 
$q_{2\bot max}^2\gg x^2 m_{\pi}^2$, we have instead
\begin{equation}
	u(z) \approx\frac{2}{3},\quad 
	 v(z)\approx z, \quad  w(z) \approx  1  .\label{uvw-asymptot}
\end{equation}
In particular we note for large values of $z$ the ratio 
$u(z)/w(z)=2/3$.

An alternative to integrating over transverse-photon momenta
is to integrate over  Compton masses $\sqrt{s}$ in the domain
$m_{\pi}^2 \leq s \leq s_{max}$. This is simply achieved by
making the integration limit $q_{2\bot max}$ depend on $s_{max}$.
 According to  Eq.(\ref{x-max-transverse}), or (\ref{q2-def}), 
 the relation between $q_{2\bot max}$ and $s_{max}$ allows
 us to write
\begin{equation}
	z(x,s_{max})=\frac{q_{2\bot max}^2}{x^2m_{\pi}^2}=
	  \frac{s_{max}}{xm_{\pi}^2} ( x_{max}  -x ), 
	  \label{z-def-bis}
\end{equation}
with
\begin{equation}
	x_{max}=1- \frac{m_{\pi}^2}{s_{max}}.\label{def_x_max}
\end{equation}
At the same time as being the definition of the variable $z$, 
Eq.(\ref{z-def-bis}) displays the functional relation between 
 $q_{2\bot max}$  and $s_{max}$.

We now specialize to 	$\sqrt{s_{max}}=3.75m_{\pi}$ which is the cut 
for the threshold region employed by the COMPASS collaboration.
This implies, via Eq.(\ref{def_x_max}), a maximal $x$-value 
of $x_{max}=0.93$. In Fig.2b the three distribution functions 
$u$, $v/10$, and $w$ are plotted as functions of $x$ for 
this value of $s_{max}$. From this figure we draw the following
conclusions.

Near the upper end of the $x$-distribution the three functions 
$u$, $v$, and $w$, may be approximated by their threshold values, 
Eq.(\ref{thesh-appr}). This is a consequence of Eq.(\ref{z-def-bis}),
which demonstrates that even though the factor $s_{max}/m_{\pi}^2=14$
is large, when $x$ is sufficiently close to $x_{max}$ the 
variable  $z$ will be small. 
This observation leads to a simplification of the 
cross-section-distribution function $F(z)$ of Eq.(\ref{F-funct}),
\begin{equation}
	F(z)= u(x,s_{max})  \left[  1  +
       \frac{s_{max}}{m_{\pi}^2} 
       ( \lambda_2 -\half \lambda_1) \right]  . \label{F-mod-prim}
\end{equation}
Numerically, we expect $\lambda_2-\half\lambda_1\approx-0.0134$ 
and small, but enters multiplied by a large factor, 
$s_{max}/m_{\pi}^2$, and therefore makes an important 
contribution to the cross-section distribution.

In the mid-range $x$-region the value of $z$ is large, due to the large
value of $s_{max}/m_{\pi}^2$, and we may here 
approximate the functions $v$ and $w$ by their asymptotic 
 expressions, Eq.(\ref{uvw-asymptot}). 
This allows us to simplify the expression for the function $F(z)$,
yielding the approximate formula
\begin{equation}
	F(x,s_{max})= u(x,s_{max})\left[1
        -\frac{3s_{max}}{4m_{\pi}^2}\ x\lambda_1
        +\frac{3x}{2(1-x)}( x  \lambda_2 + \half \lambda_1 ) \right] .
           \label{F-mod-bis}
\end{equation}
The function $u(z)$, representing the Born approximation,
 must be evaluated exactly as we are looking 
for small deviations from it.
The numerical value of the coefficient multiplying the middle
$\lambda_1$ term in Eq.( \ref{F-mod-bis}) is large.
 Nevertheless, this term has essentially no effect on the 
 cross section since according to Eq.(\ref{Lamb1-thr})
  $\lambda_1$ is proportional to the 
 sum $\alpha_{\pi} +\beta_{\pi}$ which vanishes, or 
  is very small compared with $\beta_{\pi}$.
The important term is the $\lambda_2$ term. Over most of 
the $x$-range its strength is fixed by Eq.(\ref{F-mod-bis}).
But this approximation cannot be used near 
the upper end-point since it would there overestimate the 
polarizability contribution by a factor of 3/2,
as a comparison with expression  (\ref{F-mod-prim})
would demonstrate.

\begin{figure}[ht]\label{Fin-low-mass-fig}\begin{center}
\scalebox{0.35}{\includegraphics{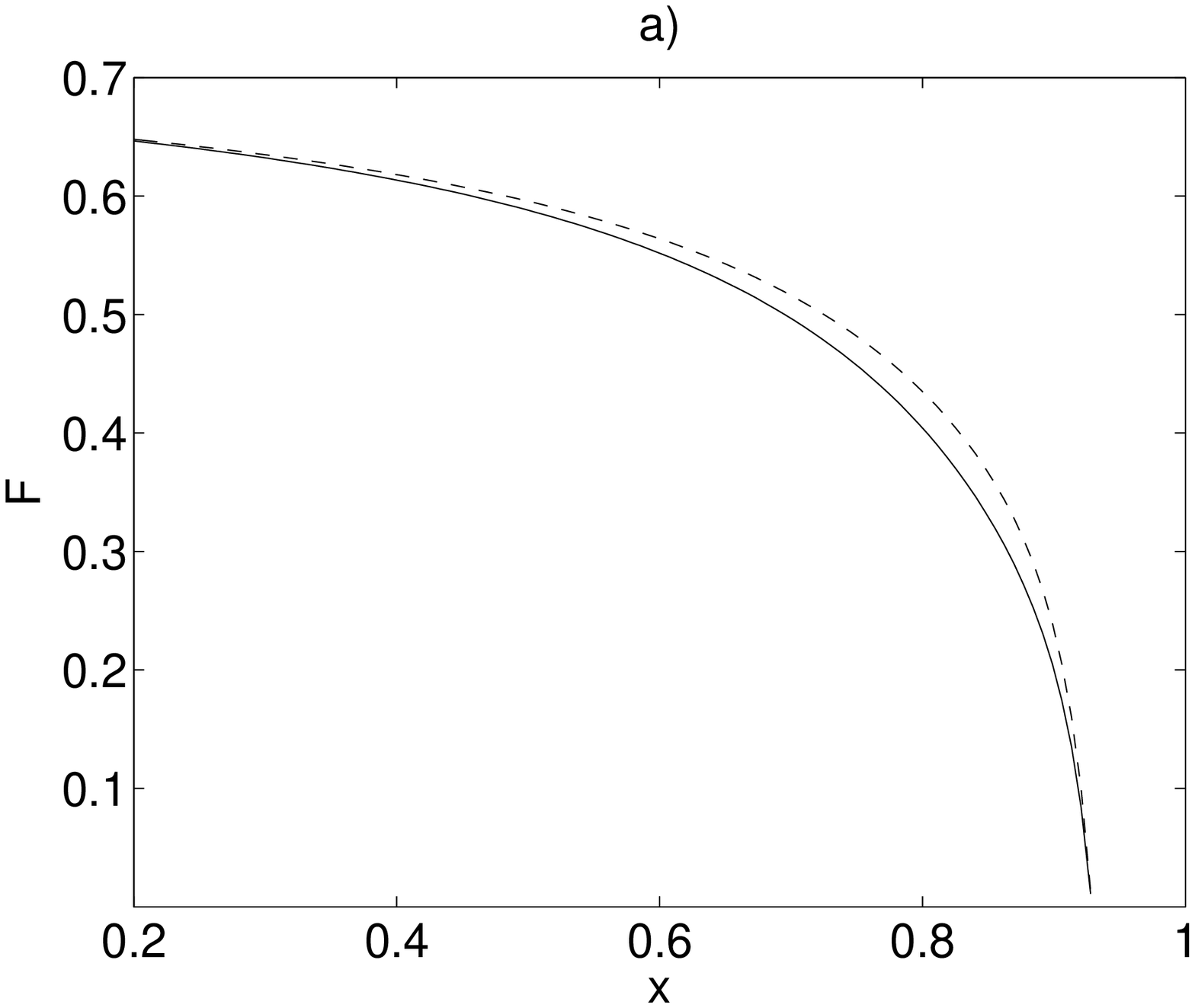} 
\quad\qquad\includegraphics{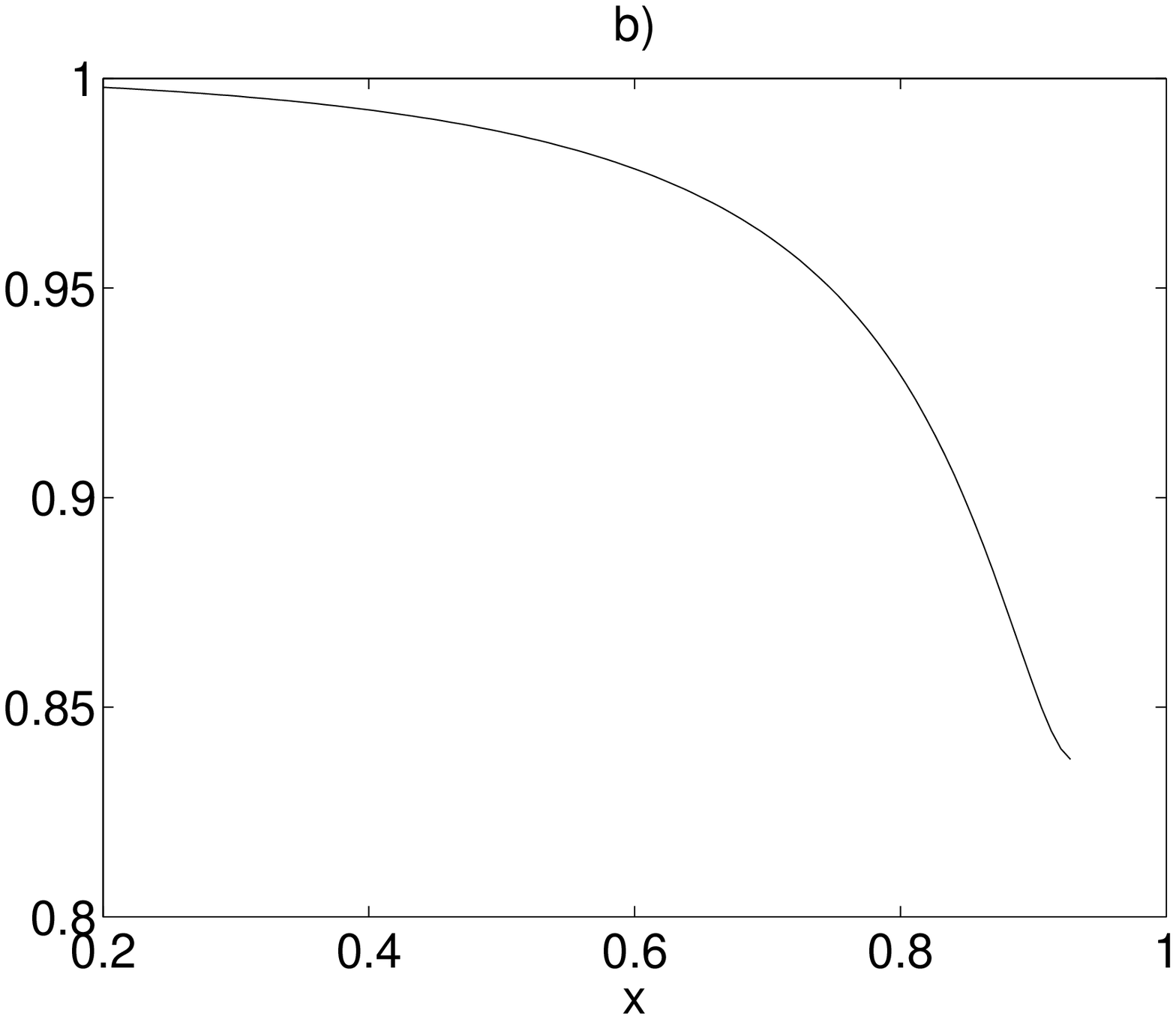}}
\end{center}
\caption{Plots of a) the distribution function $F(z(x,s_{max}))$ 
of Eq.(\ref{F-funct}) in the Born 
approximation (dashed line), and in the full calculation
 (solid line); and b) the ratio of the two distribution functions. }
\end{figure}
In Fig.3a  the 
cross-section-distribution function $F(z)$ of 
Eq.(\ref{F-funct}) is graphed in
the full calculation and in the Born approximation, 
i.e., when polarizability parameters 
$\lambda_1=\lambda_2=0$. In Fig.3b the ratio of the two
distributions is plotted. Clearly, the polarizabilities 
are important only in a region near the end-point
of the $x$- distibution, and therefore the 
experimental efforts are concentrated on this region
\cite{Ant}.

In the COMPASS experiment the apparatus is blind in a small
region around the forward direction. The transverse momenta of
the final state pions and photons must be larger than $q_{\bot cut}$
 to be detected. As a result the distribution function 
$F(z)$ of Eq.(\ref{F-funct}) must be replaced by 
\begin{equation}
	F(z)\rightarrow F(z)-F(z_0) \label{subtr}
\end{equation}
where from Eq.(\ref{z-def}) 
\begin{equation}
	z_0(x)=\frac{q_{\bot cut}^2}{x^2m_{\pi}^2}\ .
\end{equation}
In the COMPASS experiment  $q_{\bot cut}=15$ MeV/$c$ and $z_0$ is
small except when  $x$ is near $x=0$. The subtraction procedure 
of Eq.(\ref{subtr}) works unless there are momentum transfers
$q_{2\bot}< q_{\bot cut}$ for which $s> s_{max}$. This will not
happen as long as  $x$ remains in the domain
\begin{equation}
	\frac{q_{\bot cut}^2}{s_{max}-m_{\pi}^2}< x <
	    1-\frac{m_{\pi}^2}{s_{max}}-\frac{q_{\bot cut}^2}{s_{max}-m_{\pi}^2}
\end{equation}
which is the case in the COMPASS application. The effect of the cut on
the cross section is illustrated in Fig.4.
\begin{figure}[ht]\begin{center}
\scalebox{0.40}{\includegraphics{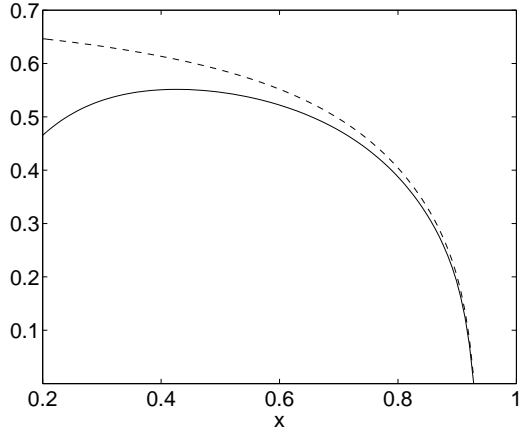}}
\end{center}
\caption{The cross-section-distribution function when small-momentum 
transfers are cut out according to Eq.(\ref{subtr}). The dashed line is 
the distribution function  without the cut, $F(z)$, and the 
solid line with the cut, $F(z)-F(z_0)$.}
\end{figure}
%
%
%
%
\clearpage
\section{Rho-mass-Compton region}

At large Compton masses the polarizability functions are 
complex and depend in a 
complicated manner on the varables $s$ and $x$.
As before, after angular integration, 
 the cross-section distribution factorizes into
one factor dependening  on $q_{1\bot}$ and $x$, and
another one depending on  $s$ and $x$. 

The integration over 
 the Coulomb-peak factor is unchanged 
 and yields the factor $P_f$ of Eq.(\ref{Peak_fact}).  As the Compton 
 energy $s$ increases and we move closer to the end-point of
 the $x$ distribution, the minimum-momentum transfer, 
 Eq.(\ref{qmin-def}), increases. 
 The maximal value is obtained at the end-point itself, which is
 located at $x=1-m_\pi^2/s$, and where $q_{min}=s/2E_1$.
At a Compton mass equals the rho mass and an energy $E_1=190$ GeV
this means $q_{min}=1.5$ MeV$/c$.
Away from $x\approx 1$ the value is much smaller.
 
The angular integration of the right hand side of 
Eq.(\ref{Coul-peak-cross-distr}) is also unchanged. However,
it is convenient to replace the variable
$q_{2\bot}^2$  by $s$, variables which are related by Eq.(\ref{Mandel}).
The cross-section distribution obtained is
\begin{equation}
\frac{\rd \sigma}{ \rd s \rd x}
  = \frac{4Z^2\alpha^3}{(s-m_{\pi}^2)^2}P_f    G(x,s) .
 \label{xs-distr}
\end{equation}
The distribution function $G(x,s)$ is given by the expresssion
\begin{equation}
	G(x,s)= 
     \left|( 1 - Y(x,s)) A(x,s)\right|^2 +
       \left| Y(x,s) A(x,s)+ B(x,s) \right|^2 , 
 \label{Pol-integrand}
\end{equation}
with $s$ related to $\mathbf{q}_{2\bot}^2$ via Eq.(\ref{Mandel}) and
\begin{equation}
	Y(x,s)=\frac{x^2m_{\pi}^2}{\mathbf{q}_{2\bot}^2+ x^2m_{\pi}^2}
	   =\frac{x}{1-x}\cdot\frac{m_{\pi}^2}{s-m_{\pi}^2}  ,
\end{equation}
with the obvious restriction $0 \leq Y \leq 1$. 

It is instructive to look at  the structure
functions as functions of $x$ and $s$. From 
Eqs (\ref{Atild-fin}) and (\ref{Btild-fin}) we derive the expressions
\begin{eqnarray}
	A(x,s)&=&  1 -\frac{1-x}{4}
	\left( \frac{s- m_{\pi}^2}{ m_{\pi}^2}\right)^2
	   \lambda_1(x, s)  , \label{Atild-fin-xs} \\
  B(x, s)&=& 
   \frac{x}{2}\left(\frac{s-m_{\pi}^2}{m_{\pi}^2}\right)
   \lambda_2(x,s)  .\label{Btild-fin-xs} 
\end{eqnarray}
 
 In the Born approximation $A=1$ and $B=0$, as 
 stated in Eqs.(\ref{ABorn}) and (\ref{BBorn}). The Born contribution,
 which is by far the dominant contribution, is in the 
 cross-section distribution (\ref{Pol-integrand}) multiplied
 by the polynomial factor
\begin{equation}
	p(Y)=1-2Y+2Y^2 , \label{pfact}
\end{equation}
 with $Y=Y(x,s)$. Since $0 \leq Y \leq 1$ it follows that
\begin{equation}
	  \frac{1}{2}\leq p(Y)\leq 1.
\end{equation}
For a fixed value of $s$, the maximal value of the pre-factor,
$p(Y)=1$, is obtained at
the boundary points $x=0$ and $x=1-m_{\pi}^2/s$. The minimum value,
$p(Y)=1/2$, is obtained at $x=1-2 m_{\pi}^2/(s+m_{\pi}^2)$. Thus, for
large values of $s/m_{\pi}^2$ there is rapid variation in the 
kinematical factor $p(Y)$ near the upper end-point of the $x$ region.
In the remaining part of  phase-space  $p(Y)$ is smoothly varying and
near to one, since there the variable $Y(x,s)$ itself will be quite small.
This is all illustrated in Fig.5.
\begin{figure}[ht]\begin{center}
\scalebox{0.60}{\includegraphics{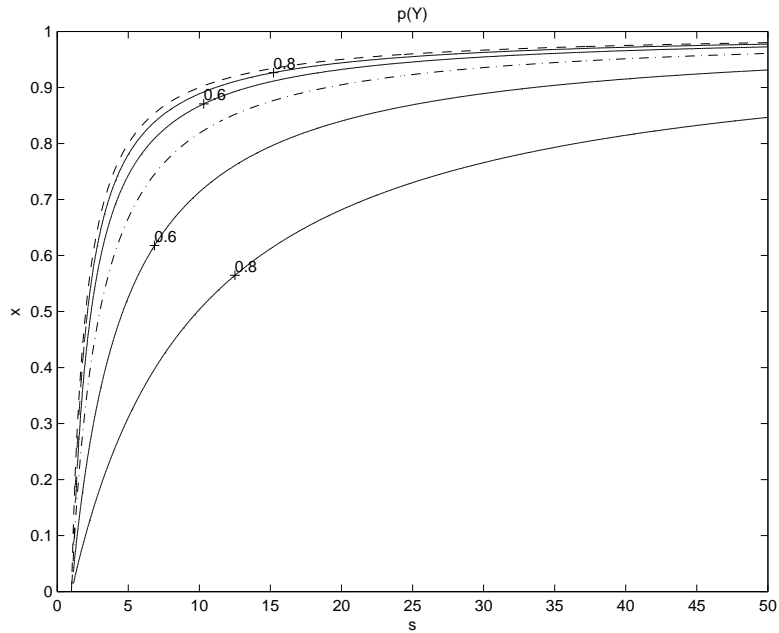}}
\end{center}
\caption{A contour plot of the kinematic factor $p(Y(x,s))$,
 or equivalently, the Born approximation to the distribution
  function $G(x,s)$. Compton mass $s$ in units of $m_{\pi}^2$. 
  The dashed line  represents the kinematic boundary,
  where $p(Y)=1$; and the dash-dotted line 
   the minimum, where $p(Y)=1/2$. }
\end{figure}

 The pion structure functions $A$ and $B$ depend on 
 the Mandelstam variables
 $s$, $t$, and $u$, of the pion-Compton scattering, or equivalently
 $x$ and $s$. In the region we are considering the Born terms dominate
 and the polarizability terms mainly come in through their
 interference with the Born terms. In Fig.6a we  plot the
  quantity $|A-1|^2$.  From Eq.(\ref{Atild-fin-xs}) it is clear that
  this is a measure of the strength of the $\lambda_1$ term.
  The graph shows that   the $\lambda_1$ term
   is small, except for $x$-values near
   $x=0$, that is grows with $s$ and exhibits a structure related
 to  the $\rho$- and $a_1$-meson exchanges.
  
  \begin{figure}[ht]\begin{center}
\scalebox{0.40}{ \includegraphics{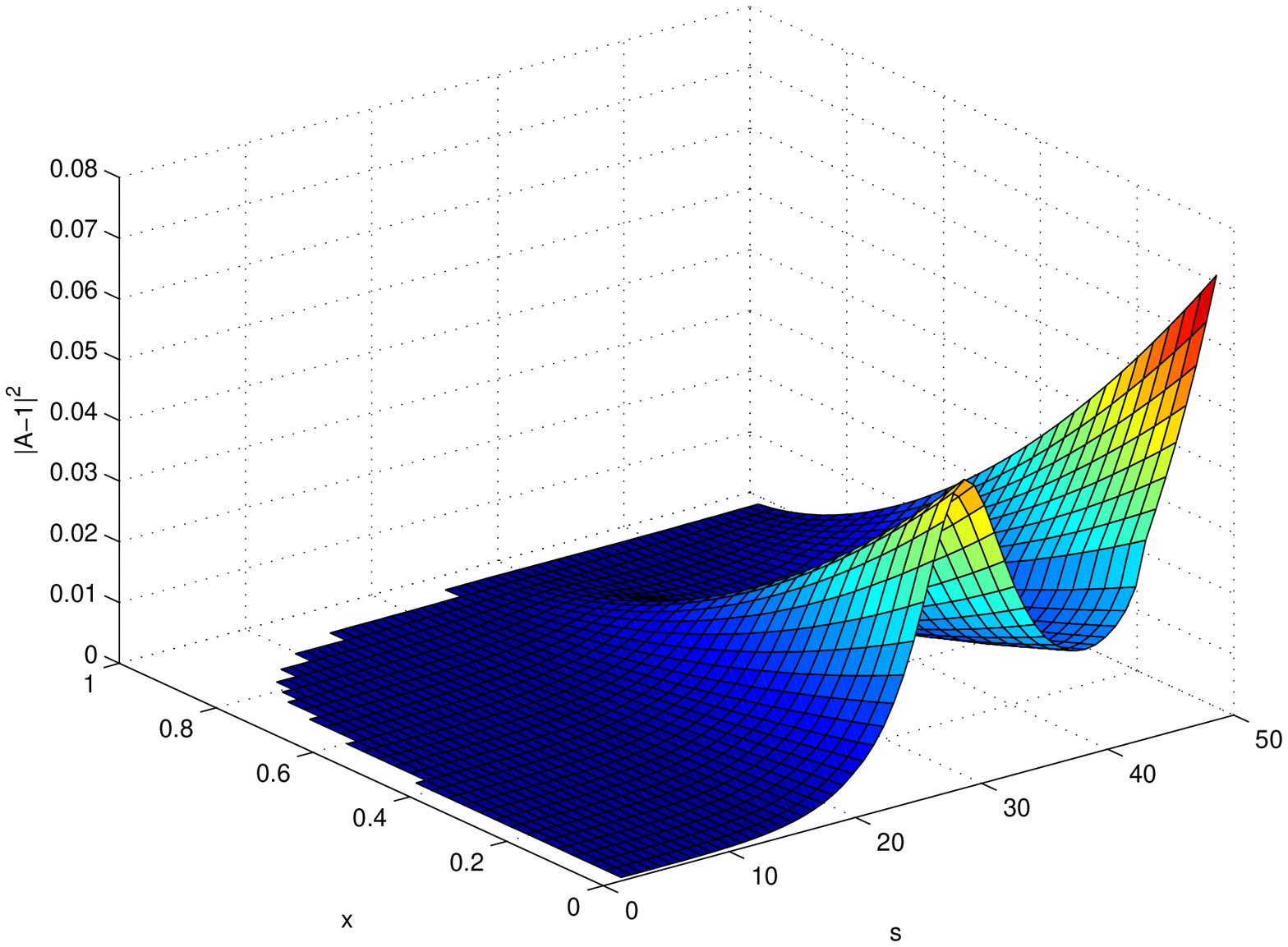} \qquad
  \includegraphics{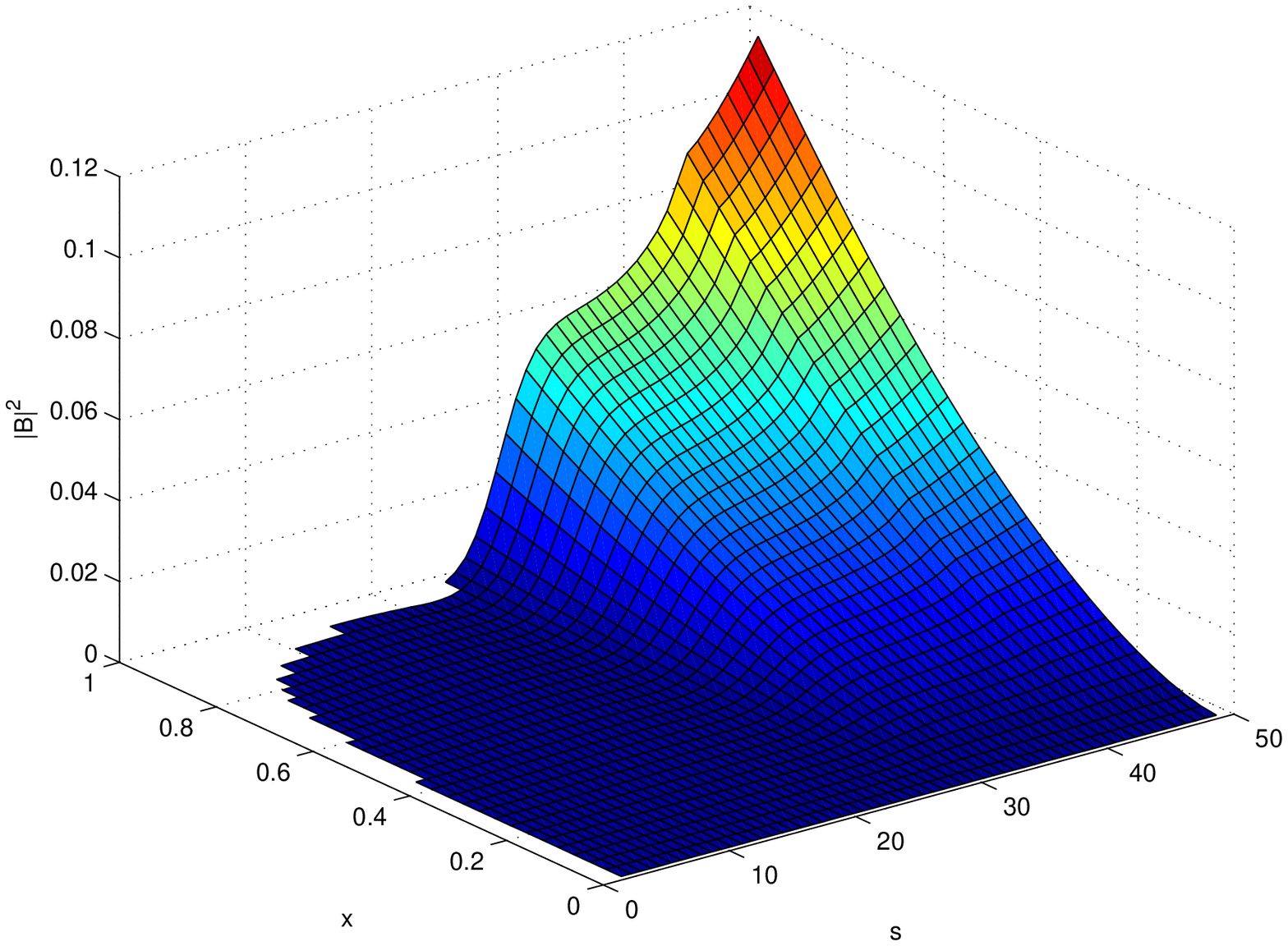}}
\end{center}
\caption{Three-dimensional plots of the functions a) $|A-1|^2$ and 
            b) $B^2$. Compton mass $s$ in units of $m_{\pi}^2$.}
\end{figure}
  In Fig.6b we plot the quantity $|B|^2$ which measures the strength 
  of the $\lambda_2$ term. The graph demonstrates that $\lambda_2$ 
  is important in a region of phase space, essentially disjoint from that 
  of $\lambda_1$,  namely the region near the upper boundary line 
  of $x$-values. 
  Its strength grows  with increasing value of $s$. 
  The structure related to the $\rho$- and $a_1$-meson exchanges is 
  visible but less pronounced. 
  It is $\lambda_2$ that contains the $\sigma$-meson exchange term. 
  According to Eq.(\ref{Pol-integrand}) the  interference between $B$ 
  and the Born term has an extra factor $Y(x,s)$. This factor results
  in a sharper structure in the interference term
   than  seen in $B$ alone. 

Fig.7 illustrates the behaviour of the dynamic factor $G(x,s)$ of the
integrand  (\ref{Pol-integrand}). The plotted quantity is the ratio
$G(x,s)/G_B(x,s)$, i.e.\ the ratio of $G(x,s)$ to its Born 
approximation $G_B(x,s)=p(Y)$.  We notice that in the overwhelming
part  of phase space the Born approximation is reasonably accurate.
The exceptions are  at small
$x$-values where the polarizability function $\lambda_1$ causes 
substantial deviations from unity, and at $x$-values near the 
upper-boundary line where the polarizability function 
$\lambda_2$ contributes to the structure.
\begin{figure}[ht]\begin{center}
\scalebox{0.60}{\includegraphics{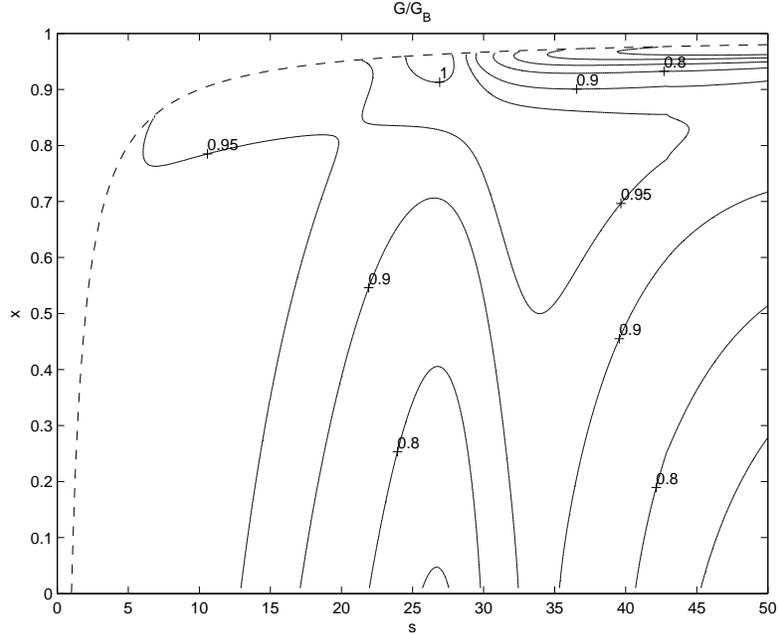}}
\end{center}
\caption{The ratio of the cross-section-distribution function 
$G(s,x)$ to its Born approximation $G_B(x,s)=p(Y)$. 
The Compton mass $s$ is measured in units of $m_{\pi}^2$, and 
the ratio $x=\omega_2/E_1$.}
\end{figure}
Unfortunately, in the latter region precise measurements are
complicated by the  sharp structure in the distribution function
$G(s,x)$.

In Fig.8a  the distribution function 
$G(x,s)$ is graphed as a function of $s$ for fixed $x=0.2$.
The dependence on the pion-polarizability functions is strong,
with a first dip at the $\rho$-meson mass. As 
pointed out above this structure is essentially caused by
the $\lambda_1$ contribution. This is seen already in Eqs.(\ref{Atild-fin-xs})
 and (\ref{Btild-fin-xs}) where the $\lambda_1$ contribution is
 multiplied by the factor $(1-x)$ and the  $\lambda_2$ contribution 
  by the factor $x$.
 \begin{figure}[ht]\begin{center}
\scalebox{0.40}{\includegraphics{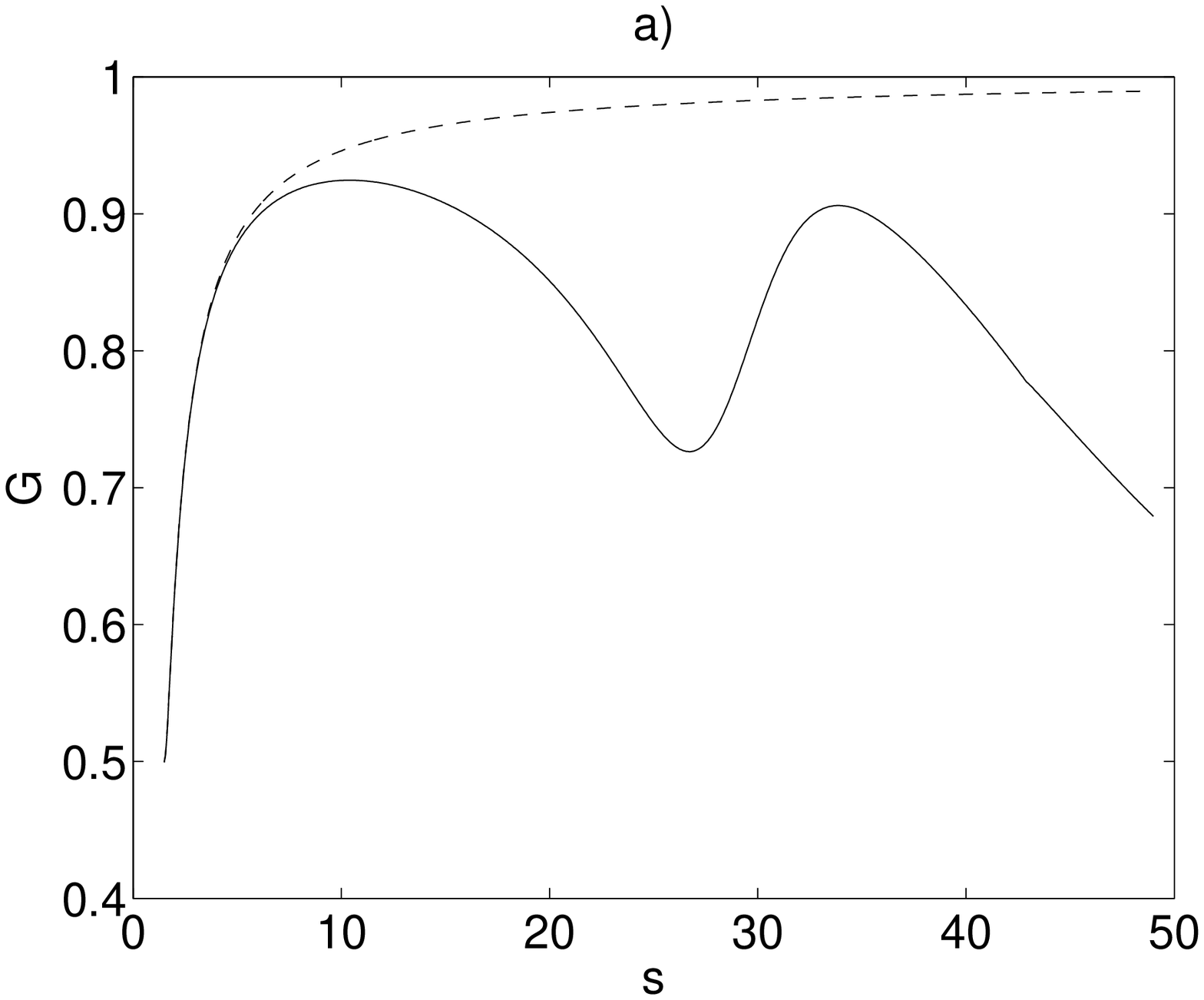}      \qquad
 \includegraphics{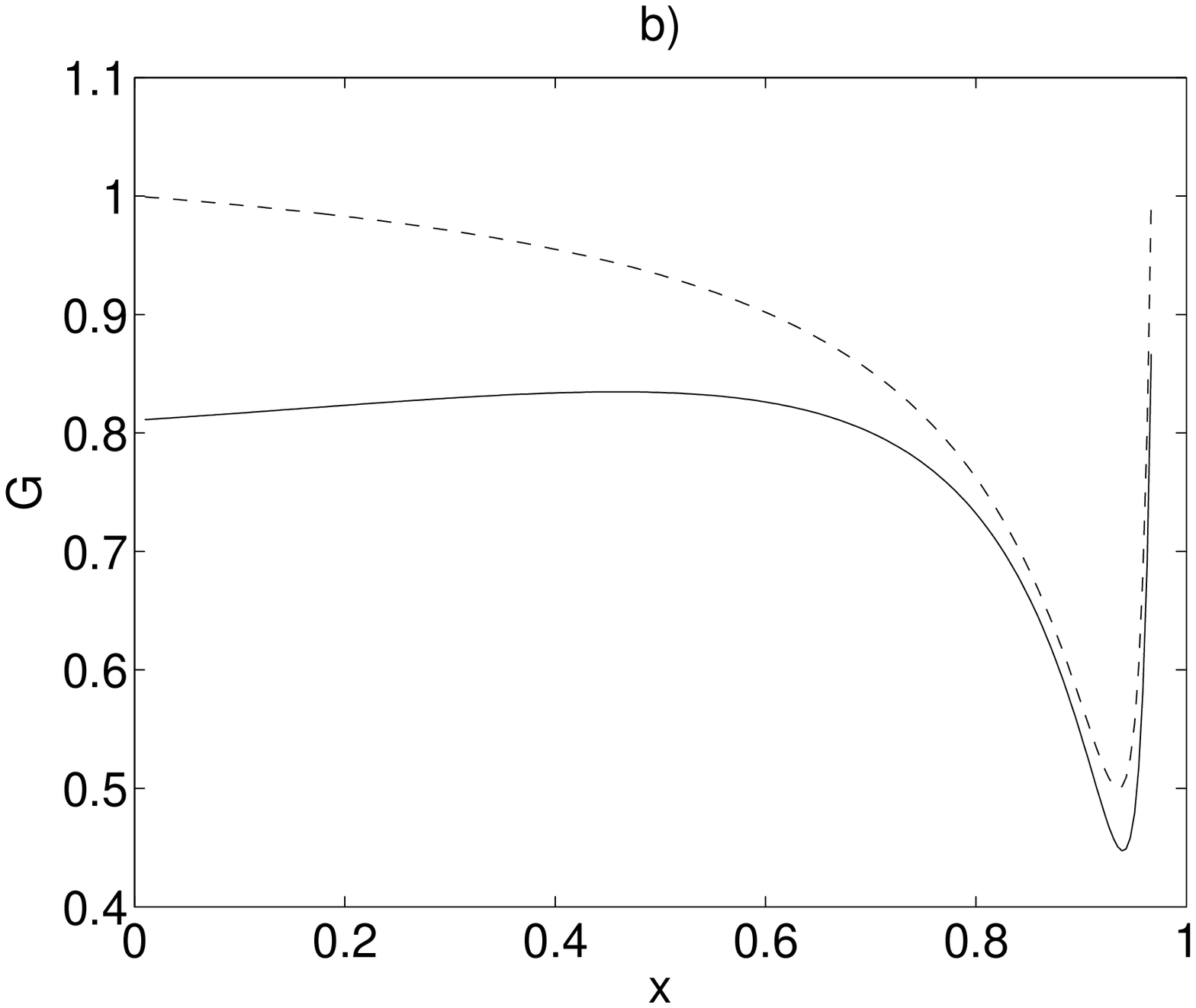}}
\end{center}
\caption{The cross-section-distribution function 
$G(s,x)$ as a) a function of  $s$, in units of $m_{\pi}^2$,
 for $x=0.2$; and 
as b) a function of $x$ for  $s=30$ $m_{\pi}^2$.}
\end{figure}
In Fig.8b the distribution function $G(x,s)$ is instead 
graphed as a function
 of $x$ for fixed $s=30$ $m_{\pi}^2$. This Compton mass is 
 right on top of the
 $\rho$-meson mass. Again we observe that the deviation from the Born
 contribution is larger for small  $x$-values  than for large
 $x$-values. Near the upper end-point there is a narrow dip, 
 a structure mainly caused be the $Y$-dependence.
%
%
\clearpage
\section{Summary}
Hard bremsstrahlung in high-energy pion-nucleus collisions 
in the Coulomb region is induced by single photon excitation.
Therefore, this reaction  can
be used to extract information about pion-Compton scattering.

The pion-Compton amplitude is dominated by Born terms
that give the exact amplitude for point-like pions.
The non-pointlike structure is described by two
independent  polarizability terms. We describe them
in a meson-exchange model with $\sigma$, $\rho$, and $a_1$, mesons.
The polarizability contributions are small but are 
important to study. At threshold in the pion-Compton system
$\lambda_1\propto(\alpha_\pi +\beta_\pi)$ is predicted to vanish 
and $\lambda_2\propto\beta_\pi$ predicted to be small. We investigate,
through the pion-bremsstrahlung mechanism, the polarizability
functions for pion-Compton c.m.\ energies $s$ below 1 GeV. 

In that part of phase space where most of the final-state 
energy is carried by the photon ($x\approx 1$), 
the contribution from $\lambda_1$
is negligible. The contribution from  $\lambda_2$
is small but its strength increases with increasing $s$.
Some structure caused by the $\rho$-and $a_1$-meson exchanges
can be seen. At the pion-Compton threshold the contribution
from the $\sigma$-exchange term dominates.
Experimental efforts to measure the polarizability 
contributions are at the moment concentrated to this region.

In that part of phase space  where most of the final-state energy 
is carried by the pion ($x\approx 1$), the contribution 
from  $\lambda_2$ is
negligible. The contribution from $\lambda_1$ is 
clearly seen, and  exhibits a strong variation due
to the   $\rho$-and $a_1$-meson exchanges. There is no contribution
from $\sigma$ exchange in $\lambda_1$.

Our analysis is valid at high energies when the transverse
momenta are small compared with the longitudinal momenta.
In addition the momentum transfer to the nucleus must be
in the Coulomb region, i.e. very small.

%
\clearpage
\section{Acknowledgements}
We would like to thank Jan Friedrich for valuable discussions
and for information about the COMPASS experiment, and Bengt Karlsson 
for help with the presentation.
\clearpage
\newpage
\section{Appendix\label{app-func}}
In Sect.3 we integrate the cross-section distribution 
Eq.(\ref{Coul-peak-cross-distr}) over the photon-transverse 
momenta $q_{2\bot}$, from the lower limit $q_{2\bot}=0$ 
up to the upper limit $q_{2\bot}=q_{2\bot max}$. 
 Then, we encounter the following
elementary integrals; for the integration of the Born contribution
\begin{eqnarray}
u(z)&=&z \int_0^{q_{2\bot max}^2} 
  \frac{ \rd q_{2\bot}^2}{q_{2\bot max}^2}
  \left( \frac{x^2 m_{\pi}^2}{\mathbf{q}_{2\bot}^2+x^2 m_{\pi}^2}\right)^2
 \left\{ 1 - \frac{2 x^2 m_{\pi}^2\mathbf{q}_{2\bot}^2}
     {(x^2 m_{\pi}^2 +\mathbf{q}_{2\bot}^2)^2} \right\} \nonumber \\
     &&\nonumber \\
&=&\frac{z}{(1+z)^2}+\frac{2z^3}{3(1+z)^3}      ;
\end{eqnarray}
for the integration of the $\lambda_1$ contribution
\begin{eqnarray}
v(z)&=&z \int_0^{q_{2\bot max}^2} 
\frac{ \rd q_{2\bot}^2}{q_{2\bot max}^2}
 \left\{ 1 - \frac{2 x^2 m_{\pi}^2\mathbf{q}_{2\bot}^2}
     {(x^2 m_{\pi}^2 +\mathbf{q}_{2\bot}^2)^2} \right\}
       \nonumber \\
        &&\nonumber \\
&=& \frac{2z+z^2}{1+z} - \ln(1+z)   ;
\end{eqnarray}
and for the integration of  the $\lambda_2$ contribution
\begin{eqnarray}
w(z)&=&z \int_0^{q_{2\bot max}^2} 
\frac{ \rd q_{2\bot}^2}{q_{2\bot max}^2}
  \left( \frac{x^2 m_{\pi}^2}{\mathbf{q}_{2\bot}^2+x^2 m_{\pi}^2}\right)
 \left\{ 1 - \frac{\mathbf{q}_{2\bot}^2}
     {x^2 m_{\pi}^2 +\mathbf{q}_{2\bot}^2} \right\} \nonumber \\
       &&\nonumber \\
&=&\frac{z}{1+z}           .
\end{eqnarray}
The parameter $z$ is defined as
\begin{equation}
	z=\frac{q_{2\bot}^2}{x^2m_{\pi}^2}  .
\end{equation}

\end{document}